\documentstyle[11pt,paspconf,epsfig]{article}

\begin{document}

\title{Temporal X-ray Properties of Galactic Black Hole Candidates: 
	Observational Data and Theoretical Models}

\author{Wei Cui}
\affil{Center for Space Research, Massachusetts Institute of Technology,
	Cambridge, MA 02139}


\begin{abstract}
I review the observed temporal X-ray properties of Galactic black hole
candidates and the status of theoretical modeling of these properties. 
Like spectral X-ray properties, which has so far attracted most of 
the attention, the temporal properties, such 
as power spectral density, hard X-ray time lags, and coherence functions, 
contain information on some of the same physical properties of inverse
Comptonization processes, which are generally believed to be responsible 
for producing hard X-ray emission in black hole candidates. 
In addition, the temporal properties add new information about the size 
of the Comptonizing region and perhaps its density structure. Therefore, 
the spectral and temporal properties taken together may help break the 
degeneracy, intrinsic to most spectral studies, in determining the 
temperature and optical depth of hot electrons. Moreover, significant 
insights might be gained into the dynamics of mass accretion in black 
hole candidates from the inferred density structure of accreted gas (from 
the temporal properties). Unfortunately, until very recently little has 
been done to {\it quantitatively} model the observed temporal properties 
of black hole candidates. I will discuss the recently proposed models, 
which have been formulated in the context of thermal Comptonization, and 
compare them to the data. Emphasis will be made on simultaneous studies 
of the observed spectral and temporal properties, the ultimate approach 
to studying X-ray production mechanisms in Galactic black holes.
\end{abstract}




\section{Introduction}
Much has been discussed about the observed spectral X-ray properties
of black hole candidates (BHCs) and numerous models that have been 
developed to account for these properties, during this workshop and in
the literature. However, the spectral properties are only part of what we 
can learn from X-ray observations of these objects. BHCs are known to 
exhibit X-ray variability on a wide range of time scales from 
milliseconds to years (see, e.g., ASM/RXTE light curves on
http://space.mit.edu/XTE/ASM\_lc.html). The temporal properties contain 
information on some of the same physical properties of inverse 
Comptonization processes (Hua \& Titarchuk 1996; Cui et al. 1997;
Kazanas et al. 1997), which are believed to be responsible for
producing hard X-ray emission from black hole candidates (Thorne \&
Price 1975; Coe et al. 1976; Shapiro et al. 1976). In addition, they 
add new information about the size of Comptonizing regions (Hua \&  
Titarchuk 1996; Cui et al. 1997), and perhaps their density structures
(Kazanas et al. 1997; Hua et al. 1999; B\"{o}ttcher \& Liang 1998). 
Unfortunately, little has been done to {\it quantitatively} model the  
temporal properties of black hole candidates until very recently.
I will review, in this paper, recent progress (both observational and 
theoretical) on this subject, with a focus on short-term variabilities
($<$ hundreds of seconds) which seem common among BHCs.

\section{Broad-Band Variability}
The discovery of rapid X-ray variability in Cyg X-1 (Oda et al. 1971) 
quickly led to the speculation that such variability might be a 
signature for BHCs. We now know that similar phenomenon has also been
observed in some of the X-ray bursters that are known to contain a
weakly magnetized neutron star (van der Klis 1995 and references
therein). Nonetheless, rapid X-ray variability is universal and 
typically very strong in BHCs. 

\subsection{Shortest Time Scale}
An interesting question is what is the shortest time scale on which X-ray 
variability has been {\it measured} with high significance.
This is a very important question because the answer to it immediately
constrains the dimension of the X-ray emitting region. In practice, however,
the answer depends on how well we can determine the level of Poisson noise
due to photon counting statistics, which in turn depends on how well we 
understand the physical processes that cause ``dead time'' in the electronics
that collects and processes the data. The latter is usually poorly understood,
so empirical models are often constructed, based on simulations, to estimate
the instrumental noise level. For instance, the Poisson noise level can be 
determined to within a fraction of one percent of the true value with this 
approach for the PCA aboard RXTE (e.g., Cui et al. 1997c). No claim should, 
therefore, be made of any detection of significant power within this level
of uncertainty.

The best studied BHC is Cyg X-1. Even in this case, significant variability
has reliably been detected only up to about 200 Hz (or 5 ms; cf. van
der Klis 1995; see also Belloni et al. 1996, Cui et al. 1997a, 1997c, 
and Nowak et al. 1998a for results from recent RXTE observations). However, 
any bursts or flares of low duty cycles would be absent from power spectral 
analyses. In a time series analysis, the estimation of the significance of 
an event on a short time scale is always complicated by the effects of 
intrinsic variability on much longer time scales (Press \& Schechter 1974). 
With this in mind, careful analyses have been conducted for Cyg X-1, and 
similar conclusions have been reached: significant variability is confirmed 
by data only for time scales longer than several milliseconds (Meekins et 
al. 1974). It is, therefore, safe to conclude that rapid X-ray variability 
has only been {\it observed} in BHCs on time scales much longer than 
dynamical time scales in the immediate vicinity of central black holes 
(assuming radiation field plays a negligible role in gas dynamics).

\subsection{Correlation with Spectral States}
\begin{figure}[th]
\centerline{\epsfig{file=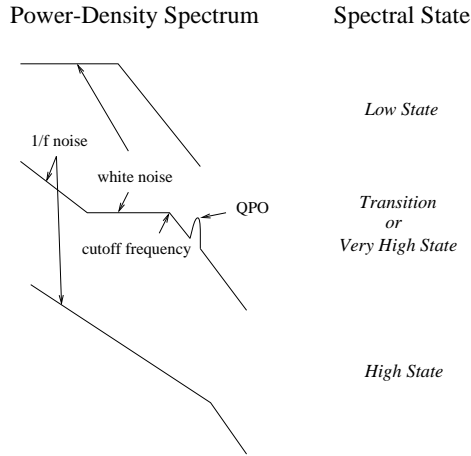,width=6cm,angle=-90}}
\vspace{-0.0in}
\caption{Canonical shapes of power spectral density (in log-log plots)
for black hole candidates in different spectral states. } \label{fig-2}
\end{figure}

For BHCs, the shape of power spectral density (PSD) correlates strongly with 
spectral states. Such correlation is illustrated in Fig.~\ref{fig-2}. 
In the low (or hard) state, the PSD can typically be characterized by a 
white noise at low frequencies that breaks into roughly a power law (with an 
index $-2$ --- $-1$) beyond certain characteristic frequency. 
The cutoff frequency, $\nu_c$, typically lies in the range 0.04--0.4 Hz. The 
level of the white noise can vary while the power-law portion remains steady. 
Belloni \& Hasinger (1990) noticed an interesting anti-correlation between 
the white noise level and $\nu_c$ for Cyg X-1: as the white noise level 
drops, $\nu_c$ increases. Various subtle features of the power-density 
spectrum have also been reported (cf. van der Klis 1995), such as another 
break at a higher frequency (a few Hz for Cyg X-1 in the low state; 
Belloni \& Hasinger 1990) which might be related to the spatial distribution
of ``seed photons'' in Comptonization processes (D. Kazanas, private 
communication).

During a transition to the high state, several things seem to happen
(Cui et al. 1997a, 1997b, 1997c): the low-frequency red noise (or the
``1/f'' noise) strengthens; the level of white noise drops; $\nu_c$ 
increases to a few Hz; and a quasi-periodic oscillation (QPO) feature 
appears. As the high state is finally reached, the PSD becomes
dominated by the 1/f noise with an apparent break at high frequencies 
(which might be similar to the second break in the low state) . It 
has also been noted that in some cases the PSD for the 
very-high state bears remarkable similarities to that for transitional 
periods, as do the spectral properties (Cui et al. 1997c; Rutledge et 
al. 1998). 

There are exceptions to the canonical correlation. Most notable are 
a subclass of BHCs that never seem exhibit high-state 
X-ray properties (Zhang et al. 1997b), even during an X-ray outburst (van 
der Klis 1995; Cui et al. 1997d), which is generally considered to be the 
high state for transient sources in terms of mass accretion rate.
The cause is still not clear (see Zhang et al. 1997b for a possible 
explanation on the spectral properties).

\subsection{Dependence on Energy and Mass Accretion Rate}
There is a general lack of systematic investigations on the energy dependence 
of broad-band X-ray variability for BHCs (see a brief summary
by van der Klis 1995). With limited statistics, the dependence seems to be 
quite weak for both the low and high states. Again use Cyg X-1 as an example, 
the recent RXTE observations of the source show that the PSD varies
little with photon energy, both in terms of the shape and the 
integrated fractional amplitude, for the low state (Nowak et al. 1998a) and 
high state (Cui et al. 1997a). If anything, the overall fractional amplitude 
appears to decrease slightly toward high energies. During the transitional
period between the two states, the energy dependence grows stronger
(Cui et al. 1997a, 1997c): the 1/f noise strengthens toward high 
energies, accompanied by a drop of the white noise level. It is worth noting 
that in this case the cutoff frequency ($\nu_c$) is rather insensitive to 
photon energy. For the very high state, the fractional amplitude seems to
increase with photon energy (Miyamoto et al. 1991; Rutledge et
al. 1998), although the energy dependence of the PSD shape seems to be 
rather complicated (Rutledge et al. 1998).

The majority of BHCs are transient X-ray sources. For these sources, spectral 
state transitions are likely to be driven by an increase in the overall rate 
of mass accretion (van der Klis 1995). Typically, the X-ray intensity is 
more variable (measured in terms of fractional amplitudes) in the low state 
(i.e., low mass accretion rate) than in the high state (van der Klis 1995).
The same is also true for Cyg X-1, a persistent source, although in this case
the bolometric X-ray luminosity (i.e., the total mass accretion rate) varies
little during the state transition (Zhang et al. 1997). The rate of mass
accretion through the disk is, however, likely to be higher in the high state,
as indicated by the sudden jump in the soft X-ray flux as the source goes from
the low state to the high state (e.g., Cui et al. 1997a; Zhang et al. 1997). 
Therefore, the state transition in Cyg X-1 is perhaps driven by a change in
the relative importance between energy dissipation processes in the disk and 
through other channels (such as the stellar wind; Zhang et al. 1997). For
several sources in the low state, there appears to be an anti-correlation
(or correlation) between the white noise level (or $\nu_c$) and the {\it disk}
accretion rate (as measured by soft X-ray fluxes) (van der Klis 1995), perhaps
implying the disk origin of noise processes.

\subsection{Hard Time Lag and Coherence Function}
It has been well established in BHCs that high-energy X-ray photons 
systematically lag low-energy photons (van der Klis 1995). Tentative 
evidence for such lags was initially found in the form of
asymmetries in cross-correlation functions between X-ray light curves in two 
energy bands for Cyg X-1 (Priedhorsky et al. 1979). At the time, the 
importance of inverse Comptonization processes was already realized
and fully appreciated: the hard power law tail on the observed energy 
spectrum for sources like Cyg X-1 was thought to be the results of
soft photons being upscattered by hot electrons in an optically thin 
region (Thorne \& Price 1975; Coe et al. 1976; Shapiro et
al. 1976). In the context of Comptonization models, the hard time lags
are expected, since more scatters are required for photons to reach
higher energies. Consequently, the observed hard lags were
conveniently interpreted this way.

To investigate hard lags for variation on different time scales, Miyamoto et 
al. (1988) computed cross power spectrum in Fourier space, using data from
Ginga observations of Cyg X-1 in the low state. As they derived the hard lag
for each Fourier frequency from the cross power spectrum, they were surprised
by the strong frequency-dependence of the measured lags, which is
entirely {\it incompatible} with the proposed 
Comptonization models that had been otherwise successful in producing the 
observed power-law energy spectrum. It can be shown that Comptonization 
processes in a uniform electron gas only produce a uniform shift between light 
curves in different energy band (Miyamoto et al. 1988); a frequency-dependent 
lag would, on the other hand, require a differential ``stretching'' of
light curves.
This result led Miyamoto et al. to question the applicability of the proposed 
models for sources like Cyg X-1. It is truly astonishing that this
extremely important observational fact seems to have been completely
forgotten or most likely ignored when models are developed to account
for the observed {\it spectral} properties. Numerous variations of 
Comptonization models have been proposed ever since. Success is often
claimed as soon as a model seem to fit the energy spectrum, without any mention
of its incompatibility with the observed frequency-dependence of hard lags;
such a model is, of course, not useful. Systematic investigations of this 
issue have just begun (Kazanas et al. 1997; Hua et al. 1999;
B\"{o}ttcher \& Liang 1998). These studies have already made us
realize that the 
incompatibility arises from the basic assumption of a uniform distribution of
Comptonizing electrons in nearly {\it all} models. Also found is an intimate 
relationship between the frequency-dependence of hard lags and the density 
distribution (spatial) of hot electrons. More specifically, the observed hard 
lags can be accounted for with the proper choice of a density profile for the 
hot electron corona. These non-uniform corona models will be discussed in more
detail in \S~4.1.

Hard time lags and their dependence on Fourier frequency appear to be a common
phenomenon for BHCs (see review by van der Klis 1995 and a summary
of recent results by Hua et al. 1999). The frequency dependence is in general
different for different sources, perhaps implying different density spatial
profiles in the context of the non-uniform corona model (Hua et al. 1999). In
fact, there are cases where no clear dependence of lags on frequency is 
detected (Miyamoto et al. 1992). Hard time lags have also been
observed for different states. There is evidence that the
frequency dependence of lags can be different in different states for
some sources (Hua et al. 1999).

\begin{figure}[th]
\centerline{\epsfig{file=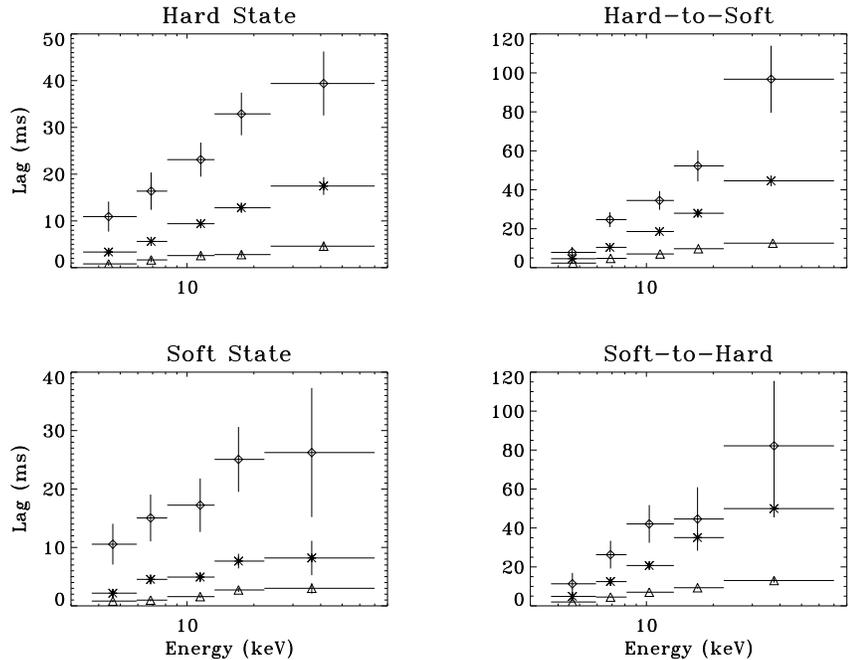,width=9cm,angle=90}}
\vspace{-0.15in}
\caption{Hard time lags with respect to 2.7 keV photons for Cyg \mbox{X-1}. 
The results for different states are plotted separately. Diamonds,
asterisks, and triangles are used to show the measured hard lags at
frequencies 0.5 Hz, 2.1 Hz, and 10 Hz, respectively.} \label{fig-6}
\end{figure}

Although the measured time lags show a general logarithmic dependence on 
photon energy (i.e., $\tau \propto log(E_h/E_l)$, where $E_h$ and $E_l$ are
the effective energies of two bands between which the lag, $\tau$, is 
measured; Cui et al. 1997c; Crary et al. 1998; Nowak et al. 1998a), as 
expected from Comptonization processes (e.g., Kazanas et al. 1997), 
the proportionality constant is dependent of Fourier frequency, 
as noted by Nowak et al. (1998a) for Cyg X-1 in the low state. I have 
carried out similar analyses for the high state and the transitional periods,
using public data from RXTE observations of Cyg X-1 during its 1996 state
transition. For completeness, I have also analyzed two public RXTE 
observations of Cyg X-1 in the low state (taken in June and October of 
1997, respectively). The results are summarized in Fig.~\ref{fig-6}. 
The energy-dependence of the slope is apparent for all states, as clearly 
shown in the figure. It is worth noting
that although the hard lag is, on average, smaller in the high state than
that in the low state the difference is not as much as previously
thought (Cui et al. 1997c). Also, the lag seems much larger during the
transitions than that in either state. This might be artificial due to
the variable nature of the low state and the randomness in our
selection of observations for this state for comparison. However, our
results for the low state are in good agreement with those of Nowak et al. 
(1998a) whose observations were taken shortly after the state transition. This 
seems to imply the stability of underlying physical processes operating in 
the low state. These results, therefore, add further complications for 
theoretical models to reconcile with (a detailed discussion of these results 
will be presented elsewhere; Cui et al. in preparation).

Complementary to hard lags, coherence functions are also of importance 
(Vaughan \& Nowak 1997). They measure the degree of linear
correlation between X-rays at different energies in Fourier
domain, normalized in practice such that a value of 1 stands for
perfect linear correlation and 0 for no correlation. If the measured
value is less than 1, the coherence function would not appear 
interesting, since numerous processes can be thought of to account 
for the loss of coherence. Observations show, however, that 
coherence functions are generally near unity for BHCs in either low or high
states (Vaughan \& Nowak 1997; Cui et al. 1997c; Smith et al. 1997;
Nowak et al. 1998a), which put useful constraints on X-ray production
processes and X-ray emitting environment (Hua et al. 1997). Moreover,
it was found that the loss of coherence is significant during
state transitions, perhaps implying the variable nature of the
Comptonizing region on a time scale less than the duration of an 
observations ($\sim$1 hour; Cui et al. 1997c).  

\section{Quasi-Periodic Oscillation}
\subsection{Phenomenology}
A QPO manifests itself in a well-localized broad peak in the PSD. The 
fractional rms amplitude is customarily quoted to specify its strength, with 
respect to the average X-ray intensity. Often, QPOs represent transient 
events. While in some cases QPOs are present in a well-defined range of 
mass accretion rate (or X-ray flux) or a well-defined spectral state, their 
presence is entirely unpredictable in other cases.

\begin{figure}[th]
\centerline{\epsfig{file=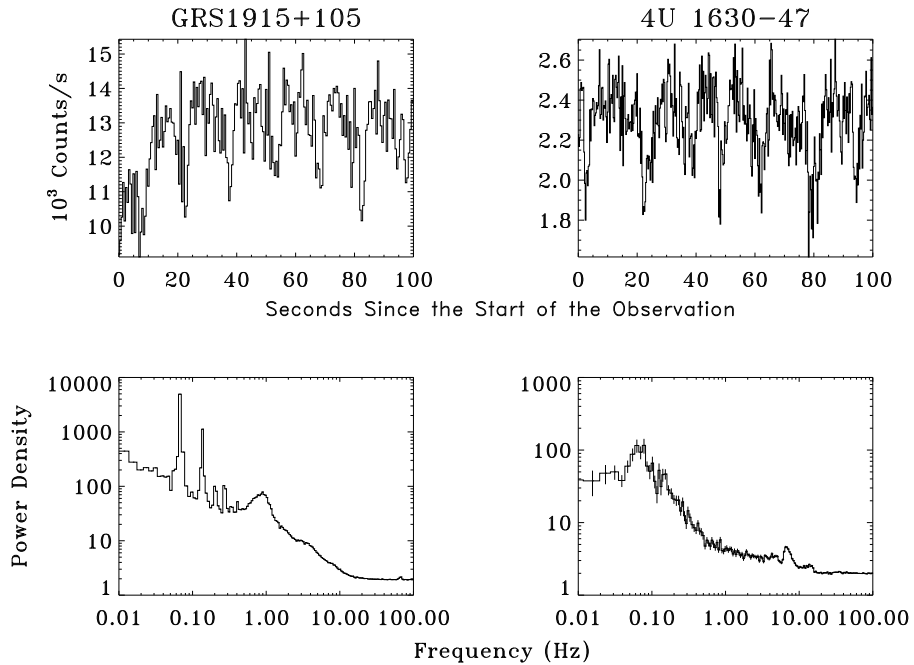,width=9cm,angle=90}}
\vspace{-0.1in}
\caption{X-ray light curves and power-density spectra for GRS 1915+105
and 4U 1630-47. The data for 4U 1630-47 are from one of the RXTE
observations during the most recent outburst in Feb, 1998. Note that
the lowest-frequency QPO corresponds to the quasi-periodic dips in the
light curve for both sources.} \label{fig-7}
\end{figure}

There is a variety of QPOs in BHCs, in terms of their basic 
characteristics (frequency, width, and strength) as well as the dependence 
of these characteristics on photon energy and X-ray intensity. As an example, 
we show in Fig.~\ref{fig-7} the power-density spectrum for GRS 1915+105 from 
a public RXTE observation made in May 5, 1996 (see also Morgan et al. 1997). 
A sharply peaked QPO is
prominent at 0.067 Hz, along with its 3 harmonics; the QPO at 0.8 Hz
is rather much broader and weaker; and even weaker is a QPO at 67
Hz. The frequency of QPOs spans three decades for this source! It
should be emphasized that not all these QPOs are present all the
time; a different set of QPOs may be revealed by a different snapshot 
(see Morgan et al. 1997 for details). GRS 1915+105 has really pushed 
the study of X-ray variability in BHCs to an extreme. Along with its cousin, 
GRO J1655-40 (M\'{e}ndez et al. 1998; Remillard et al. 1998), the two 
microquasars have begun to revolutionize our view of QPOs in BHCs. 

The study of QPOs in BHCs is also one of the most important areas that RXTE 
has contributed significantly. Prior to the launch of RXTE, the QPOs 
were observed only at very low frequencies ($< 1$ Hz; van der Klis 1995 and 
references therein, although Rutledge et al. 1998 recently reported the 
detection of QPOs of higher frequencies upon a re-analysis of GINGA data), 
except for the very-high state QPOs (at a few Hz). The RXTE observations 
have now pushed the discovery space up to a few hundred Hz (Remillard et al. 
1998). Of course, rich QPO phenomena are not limited to microquasars; they 
have also been observed in ``normal'' BHCs, such as Cyg X-1 (Cui et al. 1996, 
1997c), GRS 1758-258 and 1E 17040.7-2942 (Smith et al. 1997), 4U 1630-47 
(Dieters et al. 1998), and newly-discovered BHCs XTE J1748-2848 and
XTE J1550-564 (Cui et al., in preparation). The very low-frequency
QPOs often manifest themselves prominently in X-ray light curves,
sometimes associated with ``dipping'' activities (see Fig.~\ref{fig-7}
for examples; also Morgan et al. 1997).

\subsection{Dependence on Energy and Mass Accretion Rate}
It appears that for BHCs most QPOs strengthen (in terms of fractional 
rms amplitude) toward high energies (van der Klis 1995; Cui et al. 1997c; 
Morgan et al. 1997; Remillard et al. 1998), although some follow the
opposite trend (Morgan et al. 1997; see also Remillard et al. 1998 and
Rutledge et al. 1998 for other possible cases). In some cases, there
is a correlation between the QPO properties and the hardness of X-ray
emission (Cui et al. 1997c; Remillard et al. 1998). These facts might
be the key to understanding the origin(s) of the QPOs in BHCs, as will 
be discussed in \S~4.2.

Investigations of the dependence of QPOs on X-ray intensity are often
complicated by the ambiguity in ``labeling'' and tracking a particular
QPO over a wide intensity range. With very limited statistics, we seem
to have already seen all possibilities: some QPOs show no correlation 
with X-ray fluxes (Cui et al. 1997c); the frequency of some increases 
with X-ray fluxes, while that of others decreases (Remillard et al. 1998;
Rutledge et al. 1998); and, finally, certain QPOs seem to maintain a
stable frequency as X-ray fluxes vary (Morgan et al. 1997; Remillard
et al. 1998), i.e., the so-called ``stable QPOs'' (Cui et al. 1998a). 
Therefore, it is almost certain that no single process can account for 
all QPOs seen in BHCs. 

\subsection{Hard Time Lag}
\begin{figure}[th]
\centerline{\epsfig{file=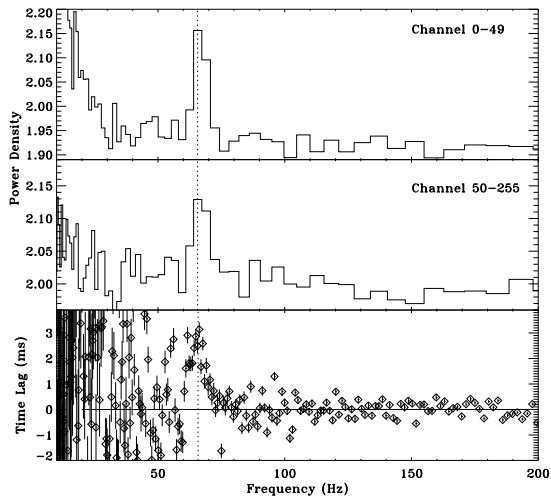,width=8cm}}
\vspace{-0.1in}
\caption{Power spectra density and hard-lag spectrum of GRS 1915+105.
The dotted-line indicates the location of the 67 Hz QPO. The energies
of the two bands are roughly 2-18.5 keV and 18.5-60 keV.} \label{fig-8}
\end{figure}

\begin{figure}[th]
\centerline{\epsfig{file=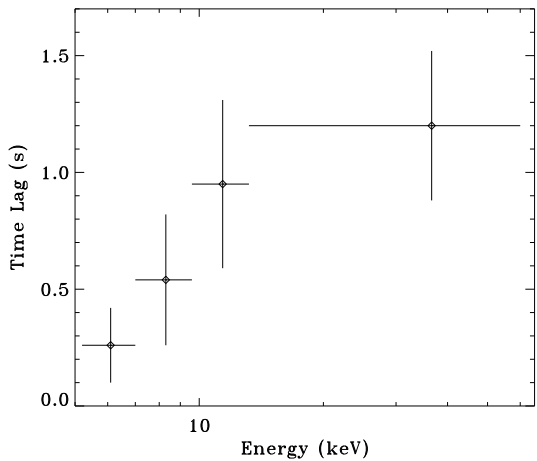,width=8cm}}
\vspace{-0.2in}
\caption{Energy dependence of QPO lags. The results shown are for the
0.133 Hz QPO in GRS 1915+105.} \label{fig-9}
\end{figure}

Like the broad-band variability, QPOs also exhibit hard time lags (van der 
Klis 1995; Cui et al. 1997c). A systematic study of QPO lags has become 
possible only recently with the RXTE discovery of a variety of QPOs in BHCs. 
For illustration, Fig.~\ref{fig-8} shows the measured hard lag of the 67
Hz QPO in GRS 1915+105 (see Fig.~\ref{fig-7}).  
At the QPO position, the time lag peaks at $\sim$3 ms. The hard lags 
have also been measured for the low-frequency QPOs in GRS 1915+105. 
Fig.~\ref{fig-9} shows the measured lags as a function of photon
energy for the 0.133 Hz QPO (see Fig.~\ref{fig-7}). 
It is clear that in one source the time lag associated with one QPO can be 
much larger than that with the others. In this case, the time lag scales 
roughly logarithmicly with photon energy, as would be expected from 
Comptonization processes. More complicated energy dependence of QPO lags 
have also been reported (van der Klis 1995). It appears to be generally 
true, however, that the larger the separation between two energy bands, 
the more hard X-ray photons lag soft ones. 

\section{Models}
\subsection{Broad-Band Noise}
Although shot-noise models (e.g., Terrell 1972; Sutherland et al. 1978) 
have been popular in describing broad-band noise properties, they have
shed little light on the physical origin(s) of underlying noise
processes. The strength of such models lies in the fact that they can 
mimic {\it mathematically} the observed light curves for
BHCs. However, it has been noted that any stochastic time series can
be simulated by a shot noise, with the proper choice of the shot
profile, up to second-order statistics such as PSD, auto- and 
cross-correlation functions (Doi 1978). Often, a distribution of shot 
profiles are required to account for subtle details (Canizares \& Oda 
1977; Ogawara et al. 1977; Nolan et al. 1981; Meekins et al. 1984; 
Belloni \& Hasinger 1990; Lochner et al. 1991; Negoro et al. 1994). 
For instance, the PSD of Cyg X-1 in the low state can drop rapidly 
above the cutoff frequency, following roughly a power law; and the 
power-law index varies in the range $-1$ to $-2$. Since the shots with a 
profile of instantaneous rise and exponential decay can only account for 
a power-law index $-2$, a distribution of lifetimes are often required 
to explain the slower drop-off. Then, a mechanism would be required to 
filter out shots with long or short lifetimes outside of the desired range, in
order to model one observed PSD; a different range would be needed to
model a different PSD, to an extreme that in some cases (e.g, Cui et
al. 1997a, 1997c) only shots with one particular lifetime would survive the
filtering process. This illustrates the ad-hoc nature of the models. 
Interesting suggestions have been made on the nature of shot-filtering
mechanisms, such as a possible application of self-organized criticality 
(Takeuchi et al. 1995), but little physics has so far been
included along this line of research. Other models employing different
statistical techniques of analyzing stochastic time series suffer from
similar drawbacks (e.g., Pottschmidt et al. 1998). Progress has also
been made recently on shot-production mechanisms (Mantomo et
al. 1998), in the context of ``advection-dominated accretion flow'' 
models (e.g, Narayan \& Yi 1995).

Since the 1/f noise is ubiquitous in nature, it should not be
surprising to see it in the process of mass accretion. For BHCs, the
noise appears to be associated with the accretion disk, because it
strengthens as the {\it disk} accretion rate increases (i.e., the high
state is approached; Cui et al. 1997a, 1997b, 1997c). One model suggests
that the low-frequency 1/f noise in BHCs might be attributed to the
fluctuation in mass accretion rate caused by the fluctuation in local
viscosity in the outer portion of the disk (Kazanas et al. 1997).

For BHCs in a luminous state, the accretion disk is likely to extend
very close to the central black hole (Tanaka \& Lewin 1995 and
references therein), where dynamical time scales are very short. 
Thermodynamic fluctuation in accretion flows in the inner region of
the disk could, therefore, manifests itself in the form of white noise
within limited frequency bands covered by current instrumentation (Cui et
al. 1997a). The disk-origin of white noise seems to be supported by
the observed anti-correlation between the relative strength of white
noise and the accretion rate through the disk (Belloni \& Hasinger
1990; also see discussion in \S~2.2). 

The power spectral break can then be naturally attributed to the
``low-pass filtering'' effects of Comptonization processes, using an
analogy in electronics, as soft (disk) photons being upscattered by
energetic electrons in the surroundings to produce the observed
hard power-law tail (Cui et al. 1997a). In the context of this model,
the cutoff frequency is determined by the characteristic escape time
of soft photons from the Comptonizing region, and thus provides a
measure of the size of such region (Cui et al. 1997a; Kazanas et
al. 1997). For Cyg X-1, the model would imply that, in the context of
thermal Comptonization models, the hot electron corona is much smaller
in the high state than in the low state, because of the much larger cutoff
frequency in the high state (Cui et al. 1997a), which is consistent with the
smaller hard lags (Cui et al. 1997c) and softer energy spectrum in the
high state, if both can also be attributed to Comptonization
processes. A smaller corona in the high state makes physical sense,
because Compton cooling power is expected to increase in this state, due 
to an increase in the supply of soft seed photons (from the disk). 

Despite early doubts from Miyamoto et al. (1988), the Comptonization
process is still a viable and likely mechanism to account for
hard lags in BHCs. It has been realized that the frequency-dependence
of the lags can be attributed to the non-uniformity of the Comptonizing
region (Kazanas et al. 1997; B\"{o}ttcher \& Liang 1998). This model
has been {\it quantitatively} applied to BHCs, and the agreement with
data can be quite good (Hua et al. 1999). As an example, while the
observed X-ray spectrum of Cyg X-1 in the high state can be fit nearly
equally well by models assuming three different density profiles 
(uniform, $\propto 1/$r, and $\propto 1/r^{3/2}$) for
the Comptonizing corona, which shows the degeneracy intrinsic to usual
spectral analyses, the observed frequency-dependence of the hard lags 
can be fit only by assuming the $1/r$ density profile (Hua et
al. 1998). To my best knowledge, this study represents the first
attempt to {\it simultaneously} model the spectral and temporal
properties of a BHC. The main drawback of such models is the 
requirement of maintaining a high temperature ($\sim$100 keV) 
throughout a large corona (roughly one light second across). It is not
clear at present whether such conditions can be satisfied physically. 

Also proposed is the possibility that the hard lags are associated
with the frequency-dependent propagation times of some waves from the
soft X-ray emitting region to the hard X-ray emitting region (Miyamoto
et al. 1988). While the physical nature of such waves are not known,
Kato (1989) suggested that low-frequency corrugation waves in a thin
accretion disk might be a possibility. However, it seems nearly 
impossible for such models to account for the lags associated with 
very high-energy photons (e.g., Grove et al. 1998). This difficulty 
can be alleviated by noting that the inner portion of the thin disk 
may be ``evaporated'' to form a geometrically 
thick, optically thin hot ``corona'' (see, e.g, Narayan \& Yi 1995), 
which can produce high-energy photons via inverse Comptonization. 
However, in order to explain large lags at low frequencies, the 
required propagation speed seems too small to be attainable in the 
context of these models (Nowak et al. 1998b); neither is it clear 
whether near perfect coherence can be preserved by the superposition 
of randomly-phased waves.

\subsection{QPO}
While it is clear that QPOs constitute a heterogeneous class of
phenomena, their origins are hardly known. Often, QPOs are attributed
to processes in the accretion disk, but evidence for such a disk
origin is scarce and inconclusive. In some cases (Cui et al. 1997c),
the properties of QPOs seem to correlate much better with those of 
hard X-rays (perhaps of a Comptonizing corona) than those of soft 
X-rays (perhaps of a disk). It is, therefore, certainly premature 
to think of QPOs as signatures for accretion disks. With these in
mind, I will list (by no means exhaustively) in the following some 
of the ideas that have been proposed over the years (see also 
review by van der Klis 1995).

Although the low-frequency QPOs ($\ll$ 100 Hz) are too slow to be 
associated with the Keplerian motion of X-ray emitting ``blobs'' 
in the inner portion of an accretion disk, they might be caused by
some obscuring structures in the outer part of the disk (Ebisawa et
al. 1989). We would then expect to see absorption dips in X-ray light
curves that are associated with the QPOs (in practice, only for very
low-frequency QPOs). While we do see intensity dips, as shown in 
Fig.~\ref{fig-7}, the dips do not always have a harder X-ray spectrum
(due to absorption) compared to the non-dip emission (e.g., Greiner et
al. 1996). In fact, some of the dips in GRS 1915+105 are thought to be
related to the ejection of matter from the disk due to thermal-viscous
instabilities (Belloni et al. 1997). Alternatively, the low-frequency 
QPOs might be related to certain oscillation modes of long wavelength 
in the accretion disk (Kato 1990). Again, the spectral properties need 
to be taken into account to assess the applicability of such models.

The high-frequency QPOs, such as the stable QPOs in GRS 1915+105 and 
GRO J1655-40, can conceivably be associated with the Keplarian motion 
at the last stable orbit around central black holes (Morgan et al. 1997), 
but at least for GRO J1655-40, such an interpretation is incompatible 
with the observed spectral properties (Cui et al. 1998b). A proposal 
has been made to associate the stable QPOs to epicyclic oscillation 
modes (Nowak et al. 1997). Not only can such modes be supported by 
accretion disks, they also become trapped in the innermost portion of 
the disk, purely due to relativistic effects (Kato \& Fukue 1980; 
Nowak \& Wagoner 1992, 1993; Perez et al. 1997). It has also been 
realized that g-mode oscillations affect the largest area of a region 
in the disk where most X-rays are emitted and are, therefore, perhaps 
most relevant to observations (Nowak et al. 1997). This model can explain 
the observed 67-Hz QPO in GRS 1915+105 and 300 Hz QPO in GRO J1655-40, 
but does not seem to be consistent with the spectral properties either,
at least for GRS 1915+105 (Cui et al. 1998b). Also proposed is that the 
stable QPOs might originate 
in the gravitomagnetic precession of accretion disks surrounding spinning 
black holes in microquasars (Cui et al. 1998a). From such a model, 
solutions have been derived to the spin of black holes in microquasars 
that are also consistent with the spectral properties (Cui et al. 1998b). 
Finally, the high-frequency QPOs might also be attributed to an 
inertial-acoustic instability in the disk (Chen \& Taam 1995). A detailed 
comparison of these models is beyond the scope of this review, and will be 
given elsewhere (Cui et al. 1998b). 

Certain QPOs might be the observational manifestation of the oscillatory 
nature of Comptonizing regions (Cui et al. 1997a; Kazanas et al. 1997; 
Titarchuk et al. 1998). Evidence lies in the correlation between the 
properties of QPOs and those of hard X-ray emission for some cases (e.g, 
Cui et al. 1997c). The oscillations might be shock-induced, and 
Comptonization processes simply take place in the post-shock region 
(Titarchuk et al. 1998).

In the very high state, the mass accretion rate is thought to be near the 
Eddington limit, so radiation drag might play a critical role in gas 
dynamics around central black holes. The QPOs observed in this state appear 
to be relatively stable in frequency (van der Klis 1995), similar to the 
normal-branch QPOs for Z sources, so they might also be caused by 
radiative-driven oscillations in a quasi-spherical accretion flow (Fortner 
et al. 1989).

Most QPOs are transient events, as already mentioned. The ever-improving
quality of observational data has begun to allow us to quantify the 
conditions in which QPOs are observed. Such information might shed
more light on the origin(s) of QPOs. Unfortunately, none of the above 
mentioned models are very explicit or quantitative about actual X-ray 
modulation processes or excitation and damping mechanisms for
oscillation or precession modes, although some makes definitive
predictions on the fundamental frequency of interest. More work is
clearly required to make progress on this subject.

\section{Outlook}
\subsection{Observation}
Clearly, our knowledge about X-ray variability in BHCs is still severely
limited by the lack of high-quality data, which is partly due to the
transient nature of many phenomena of interest. A concerted effort
among observers is urgently needed to present a complete, uniform set
of {\it reliable} results, which would include such basic properties as
light curves in different energy bands, auto- and cross-correlation 
functions, PSD, hard time lags and coherence functions as a function
of Fourier frequency, and so on, for any particular BHC at different 
X-ray fluxes (i.e., mass accretion rates) or in various spectral
states. Not only would such results allow detailed studies of the temporal
properties and their dependence on photon energy and mass accretion
rate for individual sources, they would also form a valuable database for 
carrying out 
systematic investigations of BHCs as a class, from which statistically
significant conclusions can be drawn on the common properties and
ultimately shed light on the origins of phenomena like broad-band
noise processes and QPOs. Too often, only a small subset of temporal
properties are presented for a BHC, despite the fact that more
sophisticated and complete analyses are warranted by the quality of
data. The conclusions based such limited studies can be quite
biased. For example, there is simply not sufficient evidence to support
the disk origin of {\it all} QPOs. Progress can be made by correlating
QPO properties with the characteristics of accretion disks (i.e., soft
X-rays) and Comptonizing coronae (i.e., hard X-rays). Better statistics
is also imperative for us to gain confidence in some of the
correlations that have been presented in this review.

As the quality of data is ever improving, thanks to dedicated X-ray
missions like RXTE, more emphasis should also be made on the
{\it reliability} of results. This will almost certainly require
significant effort to learn about the instruments with which data are
taken and to understand such important instrumental effects as dead 
time. For instance, as discussed in \S~2.1, the shortest time scale on
which significant X-ray variability can be detected in an observation
depends critically on our ability to estimate the Poisson noise 
level, besides other obvious factors such as the statistical quality
of the data. The estimation must take into account effects of dead
time, which are usually poorly understood. However, uncertainty can
usually be estimated based on simulations or comparison between the
model and real data. The uncertainty should always kept in mind when
results are presented. Also helpful is the presentation of enough
information so that others can evaluate the results objectively.

\subsection{Theory}
With the accumulation of high-quality data, theoretical effort should
be focused more on modeling the observed temporal X-ray properties 
in BHCs \mbox{\it quantitatively}. Desperately needed are detail models, 
like the non-uniform corona model (Kazanas et al. 1997), that make {\it
definitive} predictions about the properties of X-ray
variability. Quantities, such as light curves, auto- and
cross-correlation functions, PSD, hard lags, and coherence functions,
can then be computed from the model and be directly compared to data, 
similar to techniques in modeling spectral properties. Models
will be accepted or rejected based on the goodness of the fit. Good
models can then provide quantitative information on X-ray production
processes and environment (e.g., Hua et al. 1999).

Although certain points are worth debating, the non-uniform corona model
(Kazanas et al. 1997) represents the first attempt to describe, in a
quantitative way, the observed temporal properties in BHCs. Not only
has it overcome some of the difficulties with earlier Comptonization
models for BHCs (as discussed in \S~2.4), it also sheds light on the
dynamics of mass accretion in these sources (Kazanas et al. 1997; Hua
et al. 1998). Also attempted, for the first time, was to fit the 
observed spectral and temporal X-ray properties of Cyg X-1 in the high
state {\it simultaneously} with this model (Hua et al. 1999). Although
it was still only an eyeball fit, it was the first step in the right 
direction. Because of the degeneracy intrinsic to usual spectral
analyses, simultaneous spectral and temporal modeling represents the
ultimate approach to studying mass accretion processes in BHCs.

\acknowledgments
I would like to thank Drs. D. Kazanas and Wan Chen for informative 
discussions and helpful comments on the manuscript. It is a pleasure
to acknowledge stimulating conversations with many participants at 
the workshop. This work is supported partially by NASA through Contract 
NAS5-30612 and grants NAG5-7424 and NAG5-7484.

\end{document}